\documentclass[
reprint,
groupedaddress,
frontmatterverbose, 
bibnotes,
 amsmath,amssymb,
 aps,
physrev,
10pt,
floatfix,
]{revtex4-2}

\usepackage{graphicx}% Include figure files
\usepackage{dcolumn}% Align table columns on decimal point
\usepackage{bm}% bold math
\usepackage{color,soul}
\usepackage{comment}
\usepackage{amsmath}
\usepackage{amssymb}
\usepackage{lipsum}

\begin{document}

\preprint{APS/123-QED}

\title{``I forgot the formula:" How students can use coherence to reconstruct a (partially) forgotten equation}

\author{Katherine Gifford and Gabriel S. Ehrlich}
\affiliation{%
 Department of Physics, University of Illinois  Urbana-Champaign.}%
\email{Address all correspondence to: gifford6@illinois.edu}

\author{Engin Bumbacher}
\affiliation{%
 Haute école pédagogique du canton de Vaud}%

\author{Eric Kuo}%
\affiliation{Departments of Physics and Curriculum \& Instruction, University of Illinois Urbana-Champaign}

\date{\today}

\begin{abstract}
Introductory physics instruction emphasizes fluency with routine problem-solving procedures. However, even when applying these procedures, students frequently encounter challenges. This paper investigates how students navigate such moments when answering qualitative E\&M problems in interviews. Students frequently noted they had partially forgotten a key equation on a problem involving RC circuits. We present focal cases that show how coherence-seeking approaches were employed to overcome this problem-solving challenge. In attempts to reconstruct these equations, participants were guided by identifying and chaining qualitative dependencies and seeking coherence between qualitative and mathematical understanding of the physical system. These moments of forgetting and reconstructing equations are a useful site for studying broader physics learning goals. While prior work investigates the use of mathematical sensemaking by examining how students respond to explicit prompts, our cases illustrate how students can spontaneously use mathematical sensemaking strategies. We reflect on these cases to consider how such adaptive reasoning can be a target for instruction and assessment.
\end{abstract}

\maketitle

\section{\label{sec:level1}Introduction}

Research into student problem solving in physics has revealed a range of strategies beyond formal derivations that physics students and experts use to leverage their understanding of physical concepts to create, interpret, and check mathematical equations and solutions \cite{Hestenes1987, Sherin2001, GrecaMoreira2002, Sherin2006, Brewe2008, Uhden2012, Kuo2013, OddenRuss2018, Hahn2018, Sikorski2018, EichenlaubRedish2019, GiffordFinkel2020, Kuo2020, Zhao2021, Kuo2023}. Such strategies are categorized as \textit{mathematical sensemaking in physics}–- the different ways in which alignment between physical conceptual understanding and mathematical equations is leveraged in physics reasoning. Though an important part of disciplinary physics practice, the application of these strategies–- such as dimensional analysis, checking functional dependencies, and limiting case analysis–- is not widely evident in students' typical coursework. To increase students' use of these strategies, research commonly employs tasks that explicitly prompt students to create, interpret, or check mathematical expressions to assess these mathematical sensemaking skills \cite[e.g.,][]{Sikorski2018, Hahn2018, lenz2019surprise}.  The trade-off of this approach is that it does not reveal how strategies are opportunistically applied and coordinated with other reasoning in the course of problem solving.

 In a series of interviews where students answered qualitative E\&M problems, students commonly noted that they had (partially) forgotten a key equation on one particular problem (The Four Circuits Problem). We illustrate the range of strategies that students turn to after noting that they couldn’t precisely remember a key equation. In particular, we present focal cases that highlight the ways that students leverage coherence between their qualitative and mathematical understanding of a physical system to guide reconstruction of that equation. We explore how mathematical sensemaking relies on coherence in a particular context (between physical concepts and mathematical equations) and thus is a subset of \textit{coherence-seeking strategies} –- utilizing the inherent connection and consistency of knowledge \cite{thagard2007}. These cases illustrate students' in-situ use of mathematical sensemaking (and coherence-seeking, more broadly) to overcome a problem-solving challenge. We reflect on these cases to consider how such reasoning can be a feasible target for instruction and assessment.

\section{\label{sec:level2}Theoretical Background}
Connecting conceptual understanding to mathematical equations is core to physics and has been investigated throughout the history of physics education. Early physics education research on problem solving had a focus on (re)producing the accuracy and systematicity of expert solution methods \cite{Kuo2023}. Investigations of expert problem solving revealed that identification of the key physical concepts for a problem situation drives experts’ use of mathematics \cite{Chi1981}.  That is, the physical concepts are pointers that indicate which particular mathematical equations one should use in problem solving. These investigations have led to the development and implementation of physics problem-solving frameworks that scaffold students' initial physical analysis of the problem situation to guide the subsequent selection and use of mathematical equations \cite{Reif1982, vanheuvelen1991, heller1992, huffman1997, gaigher2007,Docktor2015, dufresne1992,larkinreif1979}. While this procedural problem-solving framework can increase students' fluency with routine procedures, this research does not explore \textit{adaptive expertise}–- the ability to productively modify existing knowledge and procedures for application in new contexts \cite{hatano1986}. Flexibility in reasoning makes one's problem-solving robust against deviations from these routine physics problem-solving procedures. Therefore, moments of forgetting an equation are a useful opportunistic research context for studying students' adaptive problem solving in physics.

What adaptive strategies could students use to reconstruct an equation that was forgotten and cannot be looked up?  One could rederive them from first principles. For instance, if one forgets the equation $v^2 = v_0^2 +2a\Delta x$, it could be rederived from the definitions of velocity and acceleration or by combining Newton's Second Law with the Work-Energy Theorem. Similarly, the time constant for RC circuits, $\tau = RC$, can be rederived by analyzing an RC circuit using Kirchhoff's Laws in combination with Ohm's law and the definition of capacitance, $Q = CV$.  

While such formal derivations are valuable, they are not the only strategy one could use. There are many ways in which an understanding of physical concepts can be used to constrain the mathematical forms that can describe those physical systems. Sherin \cite{Sherin2001} showed how undergraduate physics students could create novel equations to describe physical systems using their knowledge of symbolic forms –- conceptual schemata that are associated with particular equation symbol templates. By understanding physical quantities as being comprised of parts-of-a-whole or competing terms, students constructed equations to match those conceptual schemata without going through a first-principles derivation. For example, to describe the motion of a free-falling object that reaches terminal velocity due to air resistance, students wrote the equation $a(t) = -g + \frac{f(v)}{m}$ to indicate that the acceleration due to gravity is opposed by the acceleration due to a velocity-dependent resistance force $f(v)$.  Here, the conceptual schema of opposing influences was represented by the symbol template $\Box - \Box$, where each $\Box$ was filled in with a particular expression representing one of the two opposing influences. Investigations into symbolic forms have shown that this mathematical sensemaking strategy is applicable in a range of physics topics \cite{RyanSchermerhorn, riihiluoma2022applying, Kapodistrias2024} and can be used to explain students' problem-solving insights \cite{Kuo2013}. 

Another way mathematical sensemaking can support equation reconstruction is through the practice of checking a symbolic answer by matching its properties to the expected physical behavior. Research in this area has investigated student usage and uptake of approaches including dimensional analysis, checking functional dependencies, and limiting case analysis \cite{redish2022,EichenlaubRedish2019}. Dimensional analysis (or unit analysis) allows one to check that both sides of an equation or terms in a sum have the same dimensions (or units), a requirement for physical equations. Another method involves checking the physical plausibility of the functional dependencies within an equation–- i.e., if $X$ increases, would it make sense for $Y$ to increase, decrease, or stay the same?  One can also check that the equation produces the expected outcome for special or limiting cases.  Research has investigated how and how often students evaluate their symbolic answers in courses that explicitly teach these mathematical sensemaking strategies \cite{Hahn2018, Sikorski2018, Kuo2020}.

Mathematical sensemaking strategies are a subset of the broader scientific reasoning strategy of coherence \cite{thagard2007}, which are especially relevant in physical sciences like physics. For example, symbolic forms embody coherence by combining concepts with the symbolic structure of mathematical equations \cite{Sherin2001}. In addition to these ``implicit" uses of coherence, symbolic solution checks are explicit searches for alignment between mathematical expressions and expected physical behavior \cite{Hahn2018, Sikorski2018, Kuo2020}. Documentation of mathematical sensemaking and coherence seeking clarifies the underlying strategies and disciplinary ways of thinking that characterize physics expertise. 

 The process of chaining is another way that coherence seeking and mathematical sensemaking are operationalized. In science, \textit{chaining} is an important process, since many inferences and predictions cannot be made in one step; rather, intermediate conceptual and mathematical steps can connect the starting premises to the conclusions.  In education research, chaining processes have been studied as a central part of scientific reasoning, such as mechanistic reasoning \cite{RussHutch}, mathematical sensemaking \cite{GiffordFinkel2020}, and making qualitative inferences \cite{Lindsey2023prper,Speirs2024prper}. In the mathematical sensemaking context, chaining can provide the scaffolding necessary for mathematical and conceptual information to be linked. We will evaluate how chaining supports mathematical sensemaking and coherence-seeking practices for reasoning after forgetting an equation/dependence.

In sum, research around mathematical sensemaking in physics has investigated how physics students can (i) decide which equations to use in problem solving through conceptual analysis, (ii) create new equations from conceptual understanding, and (iii) check symbolic answers against expected physical behavior. This study builds on the existing research by investigating another activity where mathematical sensemaking is useful–- reconstructing a partially forgotten equation and more generally, reasoning after forgetting an equation. Through this work, we further elaborate the mathematical sensemaking and coherence-seeking practices that physics students can employ.

\section{\label{sec:level3}Methods}
\subsection{\label{3a}Data Collection: Interviews}
The data corpus is made up of semi-structured, one-hour, one-on-one clinical interviews involving one interviewer (EB) and a participant from one of two institutions: a community college or an elite private university. There were a total of 18 interviewees, nine from each school. From the elite private university, six students were undergraduates and three were graduate students; all were working towards degrees in physics or engineering. Students were eligible to participate if they had completed or were currently enrolled in Introductory Physics II (E\&M). All of the university students had completed this course.  

Participants were given qualitative, introductory-level E\&M questions during the interview. They were asked to reason aloud as they answered these questions. Participants were informed that this was not an evaluative activity but one focused on understanding how they thought about these questions. While remaining more neutral, the interviewer supported this framing by asking follow-up questions that were mainly for clarification of previously stated ideas. After a student completed their thinking on a question, the interviewer sometimes also introduced new ideas to see how interviewees would think about them.

\subsection{\label{3b} Analytic Approach}
A team of researchers (led by KG) produced content logs, attending to instances where physical reasoning, mathematical reasoning, and coherence between the two were evident. With this lens, we identified two problems from the interview protocol, both involving capacitors, which elicited rich reasoning across the data corpus. As we discussed these episodes and refined our interpretations, we noticed that many participants attempted to and had difficulty recalling an equation during The Four Circuits Problem (presented in the next section). ``Forgetting an equation" was an interesting site for our initial analytic focus on mathematical sensemaking, because interviewees often employed a mixture of physical and mathematical reasoning to reconstruct or compensate for the forgotten equation.

Researchers then reviewed The Four Circuits Problem in each of the 18 interviews to see how often interviewees explicitly sought to use an equation that they noted they could not exactly recall. When participants were identified as having ``forgotten an equation," their interview segment was transcribed and analyzed in detail. Through an iterative process, these analyses became more focused on particular common themes–- such as checking physical dependences, switching to a new line of reasoning, attempting to recall the equation, or displaying anxiety and uncertainty. After initial analysis, the focus on forgetting an equation expanded to include moments where students struggled to recall a variable dependence reflected within a relevant equation.  Overall, 11 students had 12 instances of either partially or totally forgetting an equation or variable dependence (one participant, Gabi, experienced two instances of "forgetting"). In the results, we present themes and illustrative examples from these cases.  Preliminary analyses of three students' reasoning have been published previously \cite{gifford2021,ehrlich2021}.

During the analytic process, we remained aware that clinical interviews do not provide unbiased access to people's knowledge and problem-solving practices. We acknowledge that the interviewee's reasoning is situated in the context of the interview, emergent from their interactions with the interviewer and framing of the interview activity. \cite{Disessa2007, Russ2012, gupta2015bridging}. Therefore, our analysis does not suggest that this is ``typical reasoning" for students that would occur outside of the interview. Nor does it claim that the interviewer's moves have no effect on what knowledge and approaches students bring to bear. Rather, our analysis describes the types of strategies students can use to navigate forgetting an equation/dependence without claiming that this is typical use or would spontaneously occur in other contexts.

\subsection{\label{3c}Focal Question: The Four Circuits Problem}
The Four Circuit Problem provides the following prompt, accompanied by the image shown in FIG. 1: 

\begin{quote}
    The light bulbs, batteries and wires are all the same in the four circuits. The capacitors in circuits A, B, C are the same. The gap in capacitor in circuit D is half of the other capacitors. All capacitors have individually been charged by connecting them to a 9V battery separately from each other. They are almost fully charged, i.e. they have the maximal possible charge on either plate.

You close the switches in all four circuits at the same time. Each capacitor takes at least 2 seconds to fully discharge. After 1 second, you open all the switches again. 

Which capacitor has the most amount of charge left on its plates? Which capacitor has the least amount of charge left on its plates? Explain your reasoning.
\end{quote}

\begin{figure}
    \centering
    \includegraphics[width=0.8\linewidth]{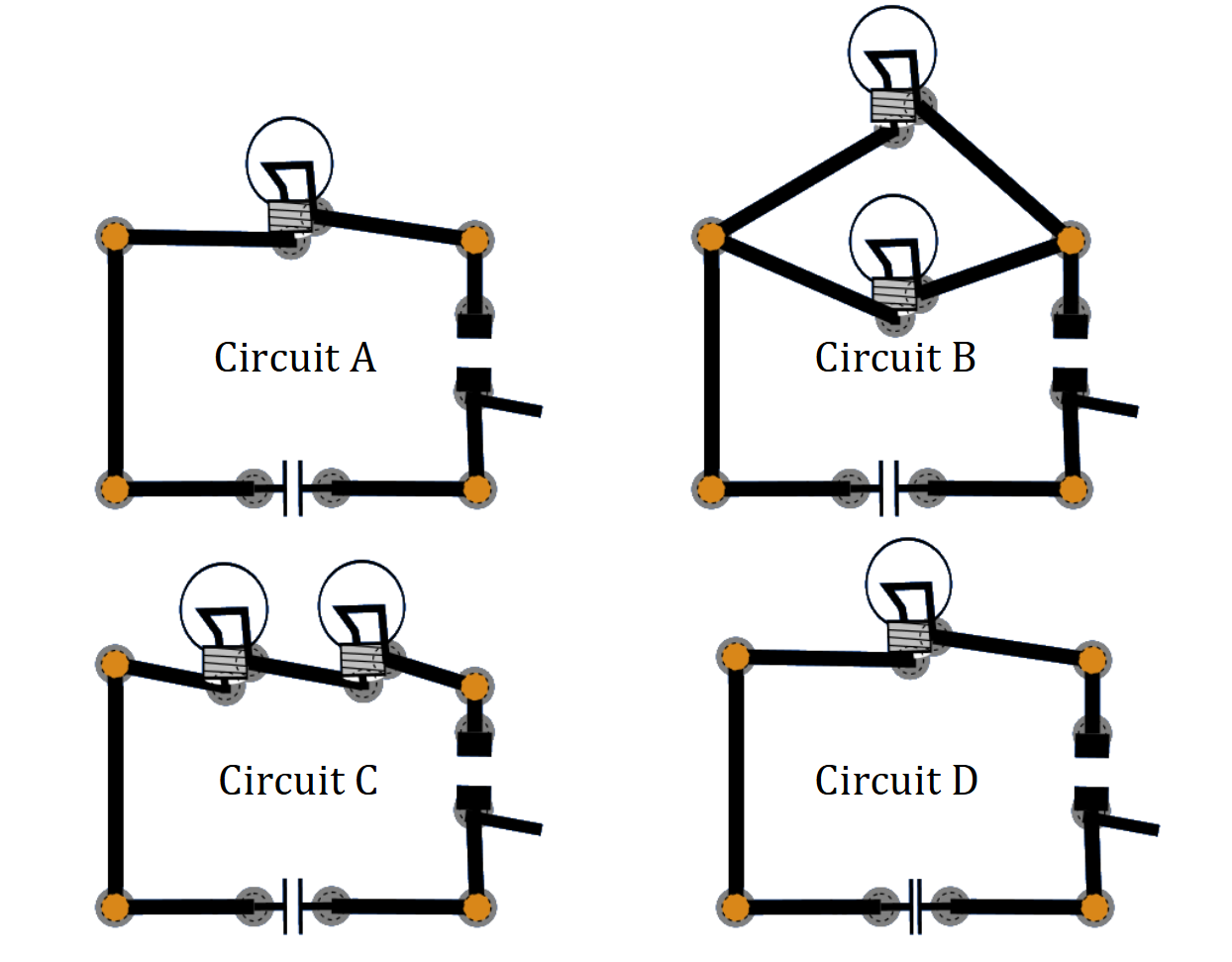}
    \caption{The Four Circuits Problem: Circuits A–-D are each shown to have a switch, ligthtbulb(s), and a capacitor. Circuits B and C have two lightbulbs, configured differently, while Circuits A and D only have one lightbulb. The gap in the capacitor in Circuit D is half the width of the gap of the other capacitors.   }
    \label{fig:FourCircuitProblem}
\end{figure}

The correct answers are that circuit D has the most amount of charge left on its plates, and circuit B has the least amount of charge left on its plates.  A symbolic method for answering this question is to compare the time constant for each RC circuit, $\tau = RC$, where $R$ is the resistance in the circuit and $C$ is the capacitance in the circuit. A larger time constant means that it will take longer for a capacitor charged to an initial voltage $V =$ 9V to discharge.  If the resistance of one light bulb is $R_0$, then the resistances of circuits A, B, C, and D are $R_0$, $R_0/2$, $2R_0$, and $R_0$, respectively.  If the capacitance in circuits A, B, and C is $C_0$, then the capacitance in circuit D is $2C_0$ since $C =  \epsilon A/d$, where $\epsilon$ is a constant, $A$ is the surface area of each capacitor plate, and $d$ is the distance between the capacitor plates. Therefore, the time constants for circuits A, B, C, and D are $R_0C_0$, $R_0C_0/2$, $2R_0C_0$, and $2R_0C_0$, respectively.  Since circuit B has the smallest time constant, it discharges the fastest and will have the least amount of charge remaining on its capacitor plates.  Circuits C and D have the largest time constant, so they take the most time to discharge.  Although circuits C and D will have the same remaining capacitor voltage, the capacitor in circuit D has more remaining charge $Q$ since it has a larger capacitance (and $Q = CV$) and was charged to a greater initial amount.  

Alternatively, one could reason qualitatively about the process of electric current and capacitor discharge to reach a similar conclusion.  However, using the equation for the time constant $\tau = RC$ allows for precise and compact comparisons of a single quantity describing discharge rate.  Therefore, recalling this time constant equation is advantageous in answering this qualitative question.  

\section{\label{sec:level4}Results}
\subsection{\label{4a}Overview}

% Table~\ref{tab:table3},%
\begin{table*}
    \caption{\label{tab:table2} Specifying the type of strategies used in the 12 cases when equations or variable dependencies were partially remembered or forgotten. The frequency of strategy usage and cases that will be presented in greater detail are noted.}%
\begin{ruledtabular}
\begin{tabular}{lcc}
\textbf{Type of Strategy Used} & \textbf{Number of Instances} & \textbf{Cases Presented}\\
\colrule
\textit{Partially Remembered Equation/Dependence}\\
\hspace{8mm}Mathematical Sensemaking to “recall” equation/dependence & 4 & \multicolumn{1}{c}{Gabi $\#$1, Tom, Gabi $\#$2, Rene}\\ 
\hspace{8mm}Spontaneous Recall & 1\\
\hspace{8mm}Selection without Reasoning & 1\\
\colrule
\textit{Totally Forgotten Equation/Dependence}\\
\hspace{8mm}Cohering Multiple Lines of Reasoning & 1 & Blake\\
\hspace{8mm}Conceptual Arguments & 2 &  Sam, Riley\\
\hspace{8mm}Interviewer Intervention & 3\\
\end{tabular}
\end{ruledtabular}
\end{table*}

As shown in Table~\ref{tab:table2}, of the 12 instances of forgetting an expression while attempting The Four Circuits Problem, six involved partially remembering an aspect of the missing expression –- students may have remembered two possible forms of an equation/dependence or may be unsure of whether a remembered equation/dependence is correct. In four of these instances, the interviewees (\textbf{Gabi $\#$1, Tom, Gabi $\#$2, and Rene}) attempted to reconstruct the missing equation or dependence using mathematical sensemaking. In the next subsections, we share illustrative cases of how mathematical sensemaking was used. The other two cases where participants partially remembered are not discussed in detail. In one of these two instances, the participant suddenly recalls the missing expression without further elaboration. In the other case, a student who recalls two possible expressions for equivalent resistance for series and parallel circuits ultimately decides one expression is for series and one expression is for parallel, without further explanation.

There were six cases where students totally forgot an equation. These students noted that there was a relevant equation, but did not discuss recalling any part of it.  The cases of \textbf{Blake, Sam, and Riley} will be shared in more detail. Blake attempts multiple lines of reasoning (conceptual arguments, reasoning qualitatively from equations, and analogical reasoning) and compares the results in order to determine the missing qualitative relationship. Two of these students, Sam and Riley, made progress by switching to purely conceptual and qualitative arguments. These students reasoned mechanistically to determine dependencies between relevant features of the Four Circuits Problem. The other three cases, which will not be discussed further, were predominately impacted by interviewer intervention to support or elicit particular reasoning pathways.

As we did not ask participants to answer demographic questions, we use they/them pronouns when referencing individual participants. Next, we present more detailed cases of these students' mathematical sensemaking and conceptual arguments.
 
\subsection{\label{4b} Mathematical Sensemaking to reconstruct equations/dependences}

\subsubsection{\label{4ci}Ex. Using qualitative dependencies to decide whether the time constant for RC circuits is $RC$ or $1/RC$}
%Gabi: Time Constant (Example of Qualitative Dependence and Chaining}

One strategy used to reconstruct partially remembered equations is connecting qualitative dependencies to the form of a mathematical equation.  \textbf{ Gabi $\#$1}, a rising junior physics major, from the elite private university, used this strategy to reconstruct the expression for the RC circuit time constant, $\tau = RC$. A preliminary analysis of Gabi's reasoning has been published previously \cite{gifford2021}.

After being presented with The Four Circuits problem, Gabi responded:

\begin{enumerate}
\item[\textbf{32}] \textbf{Gabi} Okay, so. (4s pause) The, I think that there’s like, in RC circuits like the time constant of how (5s pause), (one hand partially covers face as the other holds pen) but like the time constant of how long the capacitors take to discharge is –
\item[\textbf{33}] \textbf{Interviewer:} Mhm.
\item[\textbf{34}] \textbf{G:} $RC$?
\item[\textbf{35}] \textbf{I:} Mhm.
\item[\textbf{36}] \textbf{G:} It’s either $RC$ or $1$ over $RC$, and let’s see intuitively which one it should be. So (moves hand from head and straightens posture), as the resistance increases it would take longer, wait, wait, time constant is the amount of time it takes to go to $1$ over $e$ of its original charge.
\item[\textbf{37}] \textbf{I:} Okay.
\item[\textbf{38}] \textbf{G:} Okay. So if it was a bigger resistance, it would make sense that it would take longer, I think. And then if it was more capacitance, pshh, I have a hard time getting an intuitive grasp on like what capacitance is like (replaces hand on face), Like, so, capacitance is, is charge over, per voltage (writes “$Q/V$”), I believe.
\item[\textbf{39}] \textbf{I:} Mhm.
\item[\textbf{40}] \textbf{G:} So that’s like amount of charge that the capacitor can hold, at the, or (4s pause). So like a capacitor (moves hand from face, placing hands parallel, separated by small distance) with larger capacitance (3s pause), at the same, like, like, charge, oh here we go (points to “All capacitors have individually been charged by connecting them to a 9V battery” text), charged by the same amount, has more charge on it.
\item[\textbf{41}] \textbf{I:} Okay.
\item[\textbf{42}] \textbf{G:} So it makes sense if it has more charge on it, it would take longer, maybe sh, maybe longer to, uh, to uh, discharge. Sure, we’ll say okay, we’ll say the time constant is $RC$.

\end{enumerate}

One could mathematically derive the expression of the time constant by writing Kirchhoff’s voltage law for an RC circuit and solving a differential equation. However, Gabi used an alternative approach: determining that the time constant is $RC$, and not $1/RC$, by talking through expected qualitative dependencies of how resistance and capacitance relate to the discharge time. They stated that a larger resistance would increase the discharge time.  Although this statement was not explicitly justified with either reasoning about a physical circuit model or with relevant equations, Gabi noted that “it would make sense” (turn 38). And, while they did not
explicitly state it, the determined relationship between resistance and discharge time (along with their definition of the time constant) is sufficient information to conclude that the time constant is $RC$. Instead of making this conclusion, Gabi immediately incorporated reasoning about the other variable, capacitance. 

For capacitance, they chained together qualitative dependencies to determine how $C$ was related to discharge time.  First, in turn 38, Gabi established that capacitance is charge per voltage (following the equation $C = Q/V$). Using this, they established in turn 40 that a larger capacitance means more charge on the capacitor plates when charged to the same voltage (9V, given in the problem text). Finally, Gabi concluded that more charge means a greater discharge time (turn 42) and that therefore the time constant is $RC$. Again, Gabi stated that this conclusion ``makes sense" (turn 42) without more explicit explanation, completing a chain of proportional dependencies: ``\textit{more capacitance} $\rightarrow$ \textit{more charge} $\rightarrow$ \textit{greater discharge time}," so the time constant could go like like $C$ but not $1/C$. 

Here, we make two points about Gabi's reasoning.  First, the strategy of working out the form of the time constant $RC$ with qualitative dependencies was supported by partially recalling its form and only considering possibilities that had distinct qualitative dependencies. If Gabi had also considered possibilities like $(RC)^2$, then qualitative dependencies of ``\textit{more resistance} $\rightarrow$ \textit{greater discharge time}" would be insufficient to decide on the precise functional form.  Second, although Gabi did not provide explicit explanations for qualitative dependencies that ``made sense" to them, it is clear that they used these qualitative dependencies to determine what functional forms for the time constant are reasonable.  If we were grading Gabi's reasoning for completeness, we might deduct points for the absence of these explanations.  However, that Gabi glossed over these points to make quick progress suggests that this reasoning was not solely performative for the interviewer, where one may be more explicit in the explanations.

Gabi's reasoning that ``\textit{more capacitance} $\rightarrow$ \textit{more charge} $\rightarrow$ \textit{greater discharge time}," is an example of chaining. Specifically, to determine whether more capacitance implies a greater discharge time, Gabi used an intermediate inferential step relating to charge. This chaining provided support in determining the qualitative relationship that was then used to select the associated mathematical expressions. While it is not surprising that chaining is useful for answering a physics question \cite{RussHutch, GiffordFinkel2020, Lindsey2023prper, Speirs2024prper}, we present Gabi's reasoning to provide an example of how one can use chaining to reconstruct a partially forgotten equation.

In another example, \textbf{Tom}, a graduate student, similarly stated qualitative dependencies (without explanation) to decide whether the time constant should be RC or 1/RC:

 \begin{enumerate}
 \item[\textbf{22}] \textbf{Tom:} Um, and then, I'm trying to remember what the formula is uh–- 
 \item[\textbf{23}] \textbf{Interviewer:} For?
 \item[\textbf{24}] \textbf{T:} For, like, the time constant of an RC circuit. Um, it's either $RC$ or $1$ over $RC$. It's one of those two. Let's see if we can figure it out. If resistance, if the capac–, if we have a really, really big resistor and a really, really big capacitor, it should take a long time for it to do. And the same and the reverse is true, so it should be proportional to $RC$.
\end{enumerate}

Here, Tom used qualitative reasoning to similarly decide that the time constant for RC circuits should be $RC$ and not $1/RC$, indirectly invoking the qualitative dependences ``\textit{more resistance} $\rightarrow$ \textit{greater discharge time}" and ``\textit{more capacitance} $\rightarrow$ \textit{greater discharge time}."  Again, even though Tom did not explain why these dependencies should be true, these qualitative dependencies, rather than explicit calculation, were used to determine which partially remembered functional form was correct.

\subsubsection{\label{4cii}Ex. Chaining qualitative dependences to determine how capacitance depends on capacitor plate separation distance}
%Gabi: Capacitors (A stronger example of qualitative dependence and chaining)}

To understand the effect of capacitor plate separation, \textbf{Gabi$\#$2} again used qualitative dependencies to decide between two possible forms of a functional relationship. After Gabi concluded that the time constant is $RC$, they moved on to consider the relative resistance and capacitance of each of the four circuits. Given that the capacitor in circuit D has a different plate separation distance than the others, they considered how this would affect the capacitance. Perhaps partially recalling the equation $C = \epsilon_0 \frac{A}{d}$, Gabi worked to decide whether $C\sim A/d$ or $C\sim d/A$:

\begin{enumerate}
\item[\textbf{46}] \textbf{Gabi} Um, and the time constant for this one [circuit D], okay, so capacitance, uh also, is like area. It goes as either, like, area over distance or distance over area. If it has a larger area, it can hold more charge while being at the same voltage. 
\item[\textbf{47}] \textbf{Interviewer:} Okay.
\item[\textbf{48}] \textbf{G:} And if it has a smaller distance, it can also hold more charge at the same area. Okay, we have some fixed, wait, what am I, okay, the voltage is fixed.
\item[\textbf{49}] \textbf{I:} Okay.
\item[\textbf{50}] \textbf{G:} The voltage is fixed, so the electric field times the length, so, okay, so if the voltage is fixed [holding hands parallel to each other], you shorten the length [decrease distance between hands], it means you have to have a bigger electric field.
\item[\textbf{51}] \textbf{I:} Mhm.
\item[\textbf{52}] \textbf{G:} Bigger electric field means more charge.
\item[\textbf{53}] \textbf{I:} Mhm.
\item[\textbf{54}] \textbf{G:} So a capacitor with more, with a narrower distance has a larger capacitance.
\item[\textbf{55}] \textbf{I:} Okay.
\item[\textbf{56}] \textbf{G:} So this [circuit D] has, uh, does it say, wait, does it say how much it's –-

\item[\textbf{57}] \textbf{I:} a half, yeah.
\item[\textbf{58}] \textbf{G:} It's a half distance, so uh, it's, uh, so it [circuit D] has twice as much capacitance, so but, still one unit of resistance, so I believe it is T equals 2.

\end{enumerate}

After stating the two possible dependencies for how capacitance depends on plate area and separation, Gabi immediately stated a qualitative dependence of ``\textit{larger area} $\rightarrow$ \textit{more charge (while being at the same voltage)}." When connected with the previously invoked idea that capacitance is the amount of charge held when charged to a certain voltage (turn 40), their statement can be interpreted to mean ``\textit{larger area} $\rightarrow$ \textit{more charge} $\rightarrow$ \textit{greater capacitance}" (turn 46). As in the previous case, Gabi did not provide an explanation for this stated dependence and continued by considering the other variable, which in this case is distance.

In turn 48, Gabi started by stating a qualitative dependence, ``\textit{less plate distance} $\rightarrow$ \textit{more charge}" at the same area (turn 48) but elaborated to consider voltage more explicitly, chaining together multiple qualitative dependencies. In considering the voltage, Gabi invoked ``electric field times the length," implying the equation $V = Ed$, and then made a qualitative argument consistent with this equation: for a fixed voltage, ``\textit{less length} $\rightarrow$ \textit{bigger electric field}" (turn 50).  They then stated the dependence ``\textit{bigger electric field} $\rightarrow$ \textit{more charge}" (turn 52) and concluded that ``\textit{narrower distance} $\rightarrow$ \textit{larger capacitance} (turn 54), which is consistent with the chain of stated qualitative dependencies: for constant $V$, ``\textit{less distance} $\rightarrow$ \textit{bigger electric field} $\rightarrow$ \textit{more charge} $\rightarrow$ \textit{larger capacitance}."  The episode concluded with Gabi using the dependence $C \sim 1/d$ to determine that circuit D has twice as much capacitance as the other circuits (turn 58).

This episode once again shows how equations can be invoked for qualitative reasoning and not derivations through mathematical manipulations.  As with $C = Q/V$ in the previous episode, the equation $V = Ed$ was utilized to determine a qualitative dependence that can be incorporated into a chain of three qualitative dependencies to conclude that $C \sim 1/d$. By contrast, these equations could be combined to eliminate $V$ and find that $C = \frac{Q}{Ed}$, which is one step away from determining that $C = \epsilon_0 \frac{A}{d}$ and that $C \sim 1/d$.  The classification of Gabi's reasoning as mathematical sensemaking is precisely because they did not mathematically manipulate the equations. Instead, they consciously considered the intermediate qualitative dependencies.

Not all interviewees who attempted to chain qualitative dependencies were successful. For example, \textbf{Rene}, a graduate student, also attempted to remember how the plate separation distance affects capacitance. When prompted by the interviewer to recall other related ideas that they invoked in previous parts of the interview, Rene stated that they remembered capacitance being (directly) proportional to distance.  When asked why they think it should be (directly) proportional, Rene attempted to construct a similar chain of qualitative dependencies as Gabi did:

\begin{enumerate}
\item[\textbf{34}] \textbf{Rene:} Now, what does distance do?  I don’t remember.
\item[\textbf{35}] \textbf{Interviewer:} What do you think?  If you think about the, like, again, electrons and atoms and electric fields, and you talked before about the geometry of the plates.
\item[\textbf{36}] \textbf{R:} Yeah, yeah, yeah.  I definitely remember this, but, I remember it being proportional to $d$.
\item[\textbf{37}] \textbf{I:} Because, why do you think it should be proportional?
\item[\textbf{38}]\textbf{R:} Um [10 second pause].
\item[\textbf{39}] \textbf{I:} Like the charges are sitting on the plates, right?
\item[\textbf{40}] \textbf{R:} Yeah. [pause]  I’m trying to think of an intuitive reason for why it shoul- capacitance should be proportional to $d$.  I guess one way we can try to [inaudible] is, $Q$ is equal to $C V$ [writes $Q = CV$].  $C$ is equal to $Q$ over $V$ [writes $C = Q/V$].  [inaudible] proportional to $d$ [writes $\sim d$].  Can I make any statement that the charge should be proportional to distance? Well, that wouldn’t make much sense.  I mean, like, just the distance between them, how should that impact the amount of charge on it?  Voltage, should that be inversely proportional to $d$?  Maybe? 
\end{enumerate}  

Even though this reasoning ultimately did not provide Rene with more certainty in a particular qualitative relationship, it provides another example of how chaining qualitative dependencies is a strategy students use to justify proposed functional forms. As with Gabi, Rene related qualitative dependencies as an “intuitive” (turn 40) way of approaching the task.  The equation $C = Q/V$ was invoked again for qualitative reasoning. Here, Rene considered whether either proportionality, $Q \sim d$ or $V \sim 1/d$, made sense (turn 40) to justify $C \sim d$. 
However, unlike Gabi, they do not produce an explanation for either qualitative dependency. Gabi's recall of $V = Ed$, allowed them to make intermediate steps towards determining the relation between capacitance and distance. Without such support for these qualitative relationships, Rene was unable to conclude that these relationships``make sense."

\subsubsection{\label{4ciii} Ex. Seeking coherence between physical and mathematical lines of reasoning}

 \textbf{Blake}, an undergraduate student at the four-year university, connected multiple lines of reasoning together to answer The Four Circuits Problem.  Blake's reasoning demonstrates how students can incorporate physical explanations to justify dependences, and that finding a physical reason to believe in a mathematical explanation can help students feel more certain in an answer. Here, we overview Blake's reasoning about how the capacitor plate distance affects capacitance.  An extended analysis of Blake's reasoning has been published previously \cite{ehrlich2021}.

In considering how the capacitor plate distance affects capacitance, Blake noted that they forgot a relevant equation and then switched to reasoning about the physical motion of electrons using a line of reasoning we call ``Electrons Jumping." Blake says that, “thinking logically," having the plates closer together makes it more likely for electrons to jump across and therefore makes capacitance smaller. Although this explanation is incorrect (i.e., it is the physical mechanism of capacitor breakdown, not of ideal capacitor behavior), it leads to the dependence ``\textit{less distance} $\rightarrow$ \textit{smaller capacitance}.'' Blake does not further explain the reasoning for this dependence, but it is consistent with the reasoning that less charge means smaller capacitance. 

While Blake expressed doubt in this explanation, they proceeded with the question using this qualitative dependency.  This doubt reappears when Blake determines, using this dependence, that there is no qualitative way to decide whether circuit B or D will have more charge remaining.  While circuit B has a faster discharge rate, circuit D has less capacitance and less initial charge.  In their words, “Well, it’s like these two opposing variables ... which leads me to believe, if this conclusion [Circuit D having less capacitance] is correct at the beginning, then I wouldn’t be able to distinguish...which one has more charge."  

Subsequently, they considered a new approach, using equations to reconsider the qualitative relationship between distance and capacitance.  In this second line of reasoning, ``Capacitor Equations," Blake recalled two equations, $C=Q/V$ and $V=Ed$, and combined them to derive $C=\frac{Q}{Ed}$. Whereas Gabi chained the intermediate qualitative dependencies from each of the two equations, Blake's mathematical manipulation required the read out of only one qualitative dependence from the final equation: ``\textit{less distance} $\rightarrow$ \textit{greater capacitance}."  

At this point, Blake's mathematical approach with ``Capacitor Equations" produced a conclusion that would make the four RC circuits question solvable, though it conflicted with the physical reasoning of ``Electrons Jumping."  Following up on conclusion from ``Capacitor Equations," the interviewer asked, “Why would that be?"  While Blake could have responded with an explanation of their equation-based reasoning with ``Capacitor Equations," they instead chose to answer why it would physically make sense that ``\textit{less distance} $\rightarrow$ \textit{greater capacitance}:"  

\begin{enumerate}
\item[\textbf{35}]\textbf{I:} Why would that be? (10 s pause) Just trying to, {make sense of your reasoning}.
\item[\textbf{36}]\textbf{B:} {Right, right, right}, exactly. Um, and now I am trying to piece, like, because I do remember something about the ability for electrons to jump over, I mean that was one of the reasons, I mean, also piecing in dielectrics, the reason for improving a dielectric, adding a dielectric increases the capacitance. And what does a dielectric do? It decreases the, uh, the voltage difference between there. Um, and so having the plates closer together... wait, what? Having the plates closer together... increases the voltage difference. No wait, we just said the voltage was, um... electric field times distance… (4 s pause) Sorry, yeah yeah, having the plates closer together decreases the potential difference, if we’re assuming constant electric field…
\end{enumerate}

There are multiple indicators that Blake is seeking coherence between physical reasoning and their mathematical reasoning in ``Capacitor Equations" to answer why their reasoning makes sense.  Blake started by quickly reconsidering their physical reasoning in ``Electrons Jumping", perhaps in an attempt to "piece" together how physical lines of reasoning do or do not align with the ``Capacitor Equations" conclusion. Blake then seamlessly incorporated an alternative physical line of reasoning, ``Dielectric Analogy." Blake considers why adding a dielectric would result in an increased capacitance. By attributing this increased capacitance to the decreased the voltage difference, they were able to consider qualitative dependencies that were made implicit in ``Capacitor Equations" due to their derivation. They explicitly referenced $V = Ed$ (“...the voltage was, um... electric field times distance...") to conclude that ``\textit{less distance} $\rightarrow$ \textit{less voltage}." 

By recalling information about an analogous situation, adding a dielectric, Blake was able to reconsider the impact of decreasing the distance between capacitor plates. To be clear, while the analogy supported this chaining of ``\textit{less distance} $\rightarrow$ \textit{less voltage}$\rightarrow$ \textit{more capacitance}", Blake still relied on mathematical lines of reasoning from the equation $V = Ed$  to confirm the intermediate qualitative dependency. As Blake attempted to justify the ``Capacitor Equations" conclusion, they relied on physical analogy to chain qualitative dependencies and reasoned from equations to determine the intermediate qualitative dependency. Importantly, this reasoning aligns with the conclusion made in ``Capacitor Equations."

At the end of this episode, Blake evaluates their three lines of reasoning. Blake concluded that their “first reasoning", in ``Electrons Jumping", was wrong and that “having the plates closer together decreases the voltage difference, which increases the capacitance," a conclusion which aligns with their derived equation in ``Capacitor Equations."  However, it is not simply that Blake rejected their physical reasoning for a mathematical line of reasoning.  Rather, Blake attended explicitly to an inconsistency between their physical and mathematical reasoning and generated a new line of physical reasoning, ``Dielectric Analogy,'' that was consistent with their mathematical reasoning before concluding that it was correct.  In this way, the process of seeking coherence between physical and mathematical reasoning was part of how Blake decided which conclusions were trustworthy.

\subsection{\label{4d}Reasoning qualitatively about physical entities and processes}

In Blake's episode, qualitative reasoning from a physical mechanism initiated their reasoning. We identified two additional participants, Sam and Riley, who switched to purely qualitative reasoning about the physical entities and comparative processes in the circuits. These cases show alternative ways students reasoned after forgetting an equation that did not rely on mathematical sensemaking.

\textbf{Sam}, a community college student, explicitly noted that they ``forgot the formula" relevant to The Four Circuits Problem. A preliminary analysis of Sam's reasoning has been published previously \cite{gifford2021}. Rather than trying to reconstruct the equation, they instead switched to discussing a physical analogy of ``electrons as people" to explain the potential physical behavior of electrons. This physical analogy, which is different than the ones Blake used, has been documented before in the analogical reasoning literature \cite{GentnerGentner}. They stated, ``I treat electrons kind of like people." Sam continued, ``they kinda just make choices, you know? They kind of just pick their best-fit scenario." Sam utilized this analogy of electrons as people to reason about the physical process of current flow. They concluded that circuit B’s parallel circuit would discharge more quickly than circuit A's single path because the current ``has two different ways to go." In this way, they were able to make progress on this qualitative question without grounding their reasoning in equations.

Sam proceeded with the problem by considering the other major difference between the four circuits–- capacitor plate separation. They concluded that ``this one [D] has the most amount of charge. Because it started with the most amount of charge." Although it is not entirely clear how they determined that circuit D ends with the most charge, it may be connected to other, previously stated, qualitative relationships: the increased capacitance in circuit D due to the decreased distance and greater capacitance corresponds to greater charge. And, notably, these lines of reasoning about capacitors stemmed from another physical analogy: that capacitance indicates the ``ability [of a capacitor] to be like a battery impersonator" in that charge flows through them. 

Sam's switch to qualitative reasoning led to the correct determination of Circuit B and Circuit D having the least and most amount of charge remaining, respectively. In the absence of mathematics, they considered the physical mechanism of circuits to determine qualitative relationships between different circuit configurations and rates of discharge. Similar to Blake's third line of reasoning, a physical analogy, here ``[electrons] pick their best-fit scenario", drove this qualitative comparison.  Their other conclusion, regarding the initial amount of charge in circuit D, is plausibly connected to another analogy, capacitance as a capacitor's ability to act as a battery. While Sam does not compare the slower discharge rate of circuit C against the greater initial charge of circuit D, and there is no explicit ranking of circuit A and circuit C, Sam was able to make significant progress on this problem. In alignment with their statement, ``I forgot the formula again, but it's cool," Sam had alternative ways of reasoning that did not rely on utilizing an equation–- qualitative reasoning tied to physical analogies. 

In another example, \textbf{Riley}, an undergraduate student at the four-year university, similarly reasoned from the perspectives of electrons, articulating mechanisms involved in electron flow in the circuit to determine the relative discharge time between circuits. At the start of considering the Four Circuits Problem, they explicitly noted that they could not recall the equation that relates capacitance to plate separation distance.  Riley said, ``So there's definitely an equation, I know, that somewhere includes the distance. Do I remember it? No. [Pen held between index and middle finger, moving fingers around]." Riley proceeded with reasoning about the physical mechanisms of circuits, ``I would think that if they are closer, it is easier for the charges to move from one side to the other...Um [4 second pause], yeah, because of what's happening is that the charges are jumping from one side to the other [motions hands from right to left]." Riley added, ``So I think if they're closer together, it's easier to jump if I'm an electron (laughs)." 

 Riley began this question with the desire to use an equation, one they could not recall. Without it, Riley considered that decreased distance is qualitatively related to an increased charge movement across the plates. Following their initial difficulty in recalling an equation, Riley turned to another line of reasoning (though incorrect): their physical model of electrons jumping.  With the restated qualitative dependency, Riley’s attention returned to the prompt's question, connecting the jumping idea to the amount of remaining charge on the capacitor plates to conclude that circuit D would have the least charge remaining.

Later in the interview, Riley proceeded with ranking the circuits by attending to another feature of the circuits–- resistance. They state a qualitative relationship between increased resistance and decreased current, “\textit{greater resistance} $\rightarrow$ \textit{decreased current}." They then re-incorporate the ``electron jumping" model to generate a link between decreased current and greater remaining charge, “\textit{decreased current} $\rightarrow$ \textit{greater charge}."  The complete chain is “\textit{greater resistance} $ \rightarrow$ \textit{decreased current} $ \rightarrow$ \textit{greater charge}." This allowed Riley to conclude that circuits C and D have the most and least charge left on their plates, respectively.

Even though there are incorrect aspects to their solution, Riley demonstrated that a variety of resources are available to support problem-solving. Equations are useful, though not always available. Without the equation, Riley shifted to connect the physical model of electron jumping to chain together qualitative relationships regarding the amount of remaining charge, plate separation, and resistance.

\section{\label{5}Discussion}

\subsection{\label{5a}Coherence-seeking: an Adaptive Strategy when Forgetting an Equation}  

The results provide another empirical illustration of how students seek coherence between mathematical expressions and other ideas, in this case applied to reconstruct forgotten equations.  Though these skills are often illustrated under the label of ``mathematical sensemaking," they also speak to the practices of coherence-seeking more broadly. Gabi and Tom considered which mathematical expressions cohered with the qualitative dependences between quantities (either directly stated as ``making sense" or justified through chaining) in deciding which expression to trust. Though they were unsuccessful, Rene also used chaining to break down a proposed qualitative dependence in an attempt to justify the chained sub-dependences.  This coherence between qualitative dependences and mathematical expressions has been previously studied in the context of answer checks \cite[e.g.,][]{Hahn2018, Sikorski2018,EichenlaubRedish2019} and creating equations from one's conceptual understanding \cite{Sherin2001}. The cases in our study demonstrate how students can seek coherence with respect to mathematical sensemaking in order to navigate forgetting an equation. Specifically, we see: (a) reconstructing partially remembered equations from qualitative relationships and (b) determining qualitative relationships from remembered equations. 

Another type of coherence seeking is illustrated by Blake resolving incoherence between a physical line of reasoning (in ``Electrons Jumping") and a mathematical line of reasoning (in ``Capacitor Equations") by finding a physical reason (in ``Dielectric Analogy") that coheres with the mathematical reasoning. Although ``Dielectric analogy" does not directly relate to the physical system at hand, it is a physical example that produces the same qualitative dependence as ``Capacitor Equations," which answers why the conclusion from ``Capacitor Equations" could be believed.  This resolution between lines of reasoning resonates with previously illustrated cases of how students resolve conflicts between equation-based and conceptual approaches by finding conceptual resolutions that cohere with the math \cite{Sherin2006}. 

Coherence approaches emerge in our data when (a) qualitative relationships are determined by reasoning (i) with equations or (ii) conceptually, (b) the appropriate form of the equation is selected based on its alignment with qualitative relationships, (c) chains between qualitative dependencies are constructed, and (d) distinct lines of reasoning are evaluated against each other. These coherence-seeking strategies speak to routine and adaptive problem-solving expertise in physics education. While much of physics education focuses on developing students' fluency with the routine problem-solving procedures of the discipline, physicists adapt those routine skills in novel ways to extend our scientific understanding of the physical world \cite{Kuo2023}. 

Though the cases shown here may seem far from the innovative reasoning in physics research, they illustrate how students can innovate and adapt to a roadblock that can arise in executing even standard methods–- forgetting a relevant equation.  The reasoning presented in these cases indicates that students' problem-solving skills are not limited to brittle procedures that fail at the first sign of trouble. Here, coherence-seeking strategies like comparing equations or physical models with qualitative dependences are implemented as redundancy systems that can help students when their approach gets off track. Additionally, these strategies are consistent with the type of reasoning we'd expect expert physicists to employ. We propose that future research should investigate possibilities for developing students' adaptive expertise by teaching mathematical sensemaking and coherence-seeking skills.

\subsection{Reflections on retrieval practice}

One relevant area of research that addresses student forgetting is on retrieval practice–- the deliberate practice of attempting to recall information from memory.  Retrieval practice has been shown to improve learning and performance across multiple contexts \cite{carpenter2022science, agarwal2021retrieval}. Retrieval practice is also a growing area of interest in PER, with recent studies showing that retrieval practice can benefit factual recall and application of solution methods for physics content \cite{gjerde2020,gjerde2022problem,zu2019comparing}. 

Our results illustrate how reconstruction of partially remembered equations, namely through chaining of qualitative dependences, is an alternative to direct retrieval of physics equations from memory.  What are the implications of these results for using retrieval practice in physics instruction?  If these students had stronger recall of the RC time constant, it would have certainly been more efficient, as students would not have needed to reconstruct  $\tau = RC$ or $C \sim A/d$ to answer the Four Circuits Question.  However, prescribing retrieval practice of the RC time constant expression may discourage the use of the desirable problem-solving skills, like the coherence-seeking and mathematical sensemaking shown in our results.

That is not to say that retrieval practice could not be of service to the reconstructive processes shown.  For one, some students, like Gabi and Blake, did have good recall of other equations (e.g., $Q = CV$ and $V = Ed$). They used these recalled equations to ground their justifications for qualitative dependences. While some participants, like Tom and Gabi, appear to have confidence in certain qualitative dependences without justification, others struggle to substantiate such claims. Rene, for example, could have benefited from recalling the equation $V = Ed$.  Retrieval practice targeting recall of these (more fundamental) equations from memory can provide a reliable starting point for students' mathematical sensemaking.

Furthermore, retrieval practice that prompts looking for coherence between an equation and other ideas – including connections with other equations, physical models, or specific examples - may help students recall coherence as a skill that can be called upon in explanation, argumentation, and problem solving.  

\subsection{Connections to student epistemologies}

The reason to investigate the ability to reconstruct forgotten equations is not that it is a key learning outcome.  Because physics education values knowing how to use the equations rather than being able to recall their exact form, students often have access to equation sheets on exams and do not need to memorize the exact equations.  However, the opportunistic focus on the act of forgetting and reconstructing an equation is a useful site for studying broader physics learning goals.

For example, this work connects to physics students' epistemologies. On the Maryland Physics Expectations Survey 2 (MPEX2) \cite{McCaskey2009} –- a survey measuring students' epistemological views in physics –- one  question asks students to rate the extent to which they agree or disagree with the following statement: ``If I don't remember a particular equation needed for a problem in an exam, I can probably figure out an (ethical!) way to come up with it, given enough time.'' The ``favorable" response, reflecting experts' own beliefs and their instructional goals for students, is to agree–- to believe that physics knowledge can be reconstructed from one's knowledge rather than simply memorized (or forgotten). Some students in this study enacted these favorable epistemological beliefs by working to reconstruct a (partially) forgotten equation. Furthermore, beyond beliefs, they demonstrated the coherence-seeking practices for connecting physical concepts, qualitative dependences, and mathematical expressions that can be used to make progress when forgetting an equation.  While there has been continued work investigating the development of physics students' epistemologies, expectations, and attitudes \cite{Madsen2015}, there has been less research on the knowledge and strategies aligned with these beliefs.  Students may only believe that they can reconstruct forgotten equations if they have strategies for doing so and experiences of doing it successfully. 

\subsection{The need for assessments targeting the integration of mathematical sensemaking in practice}

The example of reconstructing equations also speaks to the challenge of detecting students' ``integrated sensemaking practices" rather than discrete mathematical sensemaking skills. Like other interview studies, we identified the ways that students use mathematical sensemaking as they answer questions and solve problems. In written assessments in a course context, one is less able to record the process of student reasoning. Therefore, it is common to assess mathematical sensemaking skills by directly asking students to demonstrate those skills either with general prompts like ``check whether this expression makes sense or not" or specific prompts eliciting specific strategies. However, this does not capture whether students would spontaneously call upon these skills in the course of their problem solving. While one study has shown that targeted free-response exam questions can detect students' spontaneous use of mathematical sensemaking in service of coherent and efficient solutions \cite{Kuo2020}, developing additional, similar assessments will be needed to facilitate mainstream adoption of mathematical sensemaking as a physics learning goal.

\section{Acknowledgements}

This material is based upon work supported by the National Science Foundation under Grant No. 1950744. Any opinions, findings, and conclusions or recommendations expressed in this material are those of the author(s) and do not necessarily reflect the views of the National Science Foundation.

\bibliographystyle{unsrt}
\bibliography{MSref2}

\begin{thebibliography}{10}

\bibitem{Hestenes1987}
D.~Hestenes.
\newblock Toward a modeling theory of physics instruction.
\newblock {\em Am. J. Phys.}, 55(5):440--454, May 1987.

\bibitem{Sherin2001}
B.~L. Sherin.
\newblock How students understand physics equations, 2001.

\bibitem{GrecaMoreira2002}
I.~M. Greca and M.~A. Moreira.
\newblock Mental, physical, and mathematical models in the teaching and learning of physics.
\newblock {\em Science Education}, 86(1):106--121, 2002.

\bibitem{Sherin2006}
B.~Sherin.
\newblock Common sense clarified: The role of intuitive knowledge in physics problem solving.
\newblock {\em Journal of Research in Science Teaching}, 43:535--555, 8 2006.

\bibitem{Brewe2008}
E.~Brewe.
\newblock Modeling theory applied: Modeling instruction in introductory physics.
\newblock {\em American Journal of Physics}, 76(12):1155--1160, 12 2008.

\bibitem{Uhden2012}
O.~Uhden, R.~Karam, M.~Pietrocola, and G.~Pospiech.
\newblock Modelling mathematical reasoning in physics education.
\newblock {\em Science and Education}, 21:485--506, 4 2012.

\bibitem{Kuo2013}
E.~Kuo, M.~M. Hull, A.~Gupta, and A.~Elby.
\newblock How students blend conceptual and formal mathematical reasoning in solving physics problems.
\newblock {\em Science Education}, 97:32--57, 1 2013.

\bibitem{OddenRuss2018}
T.~O.~B. Odden and R.~S. Russ.
\newblock Sensemaking epistemic game: A model of student sensemaking processes in introductory physics.
\newblock {\em Physical Review Physics Education Research}, 14, 11 2018.

\bibitem{Hahn2018}
K.~T. Hahn, P.~J. Emigh, M.~Lenz, and E.~Gire.
\newblock Student sense-making on homework in a sophomore mechanics course.
\newblock In {\em 2018 PHYSICS EDUCATION RESEARCH CONFERENCE}, pages 160--163. American Association of Physics Teachers (AAPT), 3 2018.

\bibitem{Sikorski2018}
T.~Sikorski, G.~D. White, and J.~Landay.
\newblock Uptake of solution checks by undergraduate physics students.
\newblock In {\em 2018 PHYSICS EDUCATION RESEARCH CONFERENCE}, pages 368--371. American Association of Physics Teachers (AAPT), 3 2018.

\bibitem{EichenlaubRedish2019}
M.~Eichenlaub and E.~F. Redish.
\newblock Blending physical knowledge with mathematical form in physics problem solving.
\newblock In {\em Mathematics in Physics Education}, pages 127--151. Springer International Publishing, 1 edition, July 2019.

\bibitem{GiffordFinkel2020}
J.~D. Gifford and N.~D. Finkelstein.
\newblock Categorical framework for mathematical sense making in physics.
\newblock {\em Physical Review Physics Education Research}, 16, 9 2020.

\bibitem{Kuo2020}
E.~Kuo, M.~M. Hull, A.~Elby, and A.~Gupta.
\newblock Calculation-concept crossover: Assessing mathematical sensemaking in physics as seeking coherence between calculations and concepts.

\bibitem{Zhao2021}
F.~F. Zhao and A.~Schuchardt.
\newblock Development of the sci-math sensemaking framework: categorizing sensemaking of mathematical equations in science, 12 2021.

\bibitem{Kuo2023}
E.~Kuo.
\newblock {\em Cognitive, Epistemic, and Affective Theoretical Underpinnings of Research on Physics Learning Two Perspectives on Physics Problem Solving and Their Relation to Adaptive Expertise}, pages 1--26.
\newblock AIP Publishing, 3 2023.

\bibitem{lenz2019surprise}
M.~Lenz, P.~J. Emigh, and E.~Gire.
\newblock Surprise! students don't do special-case analysis when unaware of it.
\newblock In {\em 2018 Physics Education Research Conference (PERC)}, pages 231--234, 2019.

\bibitem{thagard2007}
P.~Thagard.
\newblock Coherence, truth, and the development of scientific knowledge.
\newblock {\em Philosophy of science}, 74(1):28--47, 2007.

\bibitem{Chi1981}
M.~T.H. Chi, P.~J. Feltovich, and R.~Glaser.
\newblock Categorization and representation of physics problems by experts and novices.
\newblock {\em Cognitive Science}, 5:121--152, 1981.

\bibitem{Reif1982}
F.~Reif and J.~I. Heller.
\newblock Knowledge structure and problem solving in physics.
\newblock {\em Educational Psychologist}, 17:102--127, 6 1982.

\bibitem{vanheuvelen1991}
A.~Van Heuvelen.
\newblock Overview, case study physics.
\newblock {\em American Journal of Physics}, 59(10):898--907, 10 1991.

\bibitem{heller1992}
P.~Heller, R.~Keith, and S.~Anderson.
\newblock Teaching problem solving through cooperative grouping. part 1: Group versus individual problem solving.
\newblock {\em American Journal of Physics}, 60(7):627--636, 07 1992.

\bibitem{huffman1997}
D.~Huffman.
\newblock Effect of explicit problem solving instruction on high school students' problem-solving performance and conceptual understanding of physics.
\newblock {\em Journal of Research in Science Teaching}, 34(6):551--570, 1997.

\bibitem{gaigher2007}
E.~Gaigher, J.~M. Rogan, and M.~W.~H. Braun.
\newblock Exploring the development of conceptual understanding through structured problem‐solving in physics.
\newblock {\em International Journal of Science Education}, 29(9):1089--1110, 2007.

\bibitem{Docktor2015}
J.~L. Docktor, N.~E. Strand, J.~P. Mestre, and B.~H. Ross.
\newblock Conceptual problem solving in high school physics.
\newblock {\em Physical Review Special Topics - Physics Education Research}, 11, 9 2015.

\bibitem{dufresne1992}
R.~J. Dufresne, W.~J. Gerace, P.~Thibodeau Hardiman, and J.~P. Mestre.
\newblock Constraining novices to perform expertlike problem analyses: Effects on schema acquisition.
\newblock {\em Journal of the Learning Sciences}, 2(3):307--331, 1992.

\bibitem{larkinreif1979}
J.~H. Larkin and F.~Reif.
\newblock Understanding and teaching problem‐solving in physics.
\newblock {\em European Journal of Science Education}, 1(2):191--203, 1979.

\bibitem{hatano1986}
G.~Hatano and K.~Inagaki.
\newblock Two courses of expertise.
\newblock {\em Child development and education in Japan}, 17:262--272, 1986.

\bibitem{RyanSchermerhorn}
Q.~X. Ryan and B.~P. Schermerhorn.
\newblock Students' use of symbolic forms when constructing equations of boundary conditions.
\newblock {\em Phys. Rev. Phys. Educ. Res.}, 16:010122, Apr 2020.

\bibitem{riihiluoma2022applying}
W.~D. Riihiluoma, Z.~Topdemir, and J.~R. Thompson.
\newblock Applying a symbolic forms lens to probability expressions in upper-division quantum mechanics.
\newblock In {\em 2022 Physics Education Research Conference Proceedings}, pages 383--388, 2022.

\bibitem{Kapodistrias2024}
A.~Kapodistrias and J.~Airey.
\newblock Rearranging equations to develop physics reasoning.
\newblock {\em Eur. J. Phys.}, 45(3), 2024.

\bibitem{redish2022}
E.~Redish.
\newblock Using math in physics: 5. functional dependence.
\newblock {\em The Physics Teacher}, 60(1):18--21, 01 2022.

\bibitem{RussHutch}
R.~S. Russ and P.~Hutchison.
\newblock It's okay to be wrong: Recognizing mechanistic reasoning during student inquiry.

\bibitem{Lindsey2023prper}
B.~A. Lindsey, M.~R. Stetzer, J.~C. Speirs, W.~N. Ferm, and A.~van Hulten.
\newblock Investigating student ability to follow reasoning chains: The role of conceptual understanding.
\newblock {\em Phys. Rev. Phys. Educ. Res.}, 19:010128, Apr 2023.

\bibitem{Speirs2024prper}
J.~C. Speirs, M.~R. Stetzer, and B.~A. Lindsey.
\newblock Utilizing network analysis to explore student qualitative inferential reasoning chains.
\newblock {\em Phys. Rev. Phys. Educ. Res.}, 20:010147, May 2024.

\bibitem{gifford2021}
K.~Gifford, G.S. Ehrlich, E.~Bumbacher, and E.~Kuo.
\newblock Seeking coherence and switching reasoning after forgetting an equation.
\newblock In {\em 2021 Physics Education Research Conference Proceedings}, 2021.

\bibitem{ehrlich2021}
G.S. Ehrlich, K.~Gifford, E.~Kuo, and E.~Bumbacher.
\newblock Seeking physical/mathematical coherence by recruiting and reconciling reasoning: A case study in e\&m.
\newblock In {\em 2021 Physics Education Research Conference Proceedings}, 2021.

\bibitem{Disessa2007}
A.~A. Disessa.
\newblock An interactional analysis of clinical interviewing, 2007.

\bibitem{Russ2012}
R.~S. Russ, V.~R. Lee, and B.~L. Sherin.
\newblock Framing in cognitive clinical interviews about intuitive science knowledge: Dynamic student understandings of the discourse interaction.
\newblock {\em Science Education}, 96:573--599, 7 2012.

\bibitem{gupta2015bridging}
A.~Gupta, A.~Elby, and V.~Sawtelle.
\newblock Bridging knowledge analysis and interaction analysis through understanding the dynamics of knowledge in use.
\newblock In {\em Knowledge and interaction}, pages 276--307. Routledge, 2015.

\bibitem{GentnerGentner}
D.~Gentner and D.~Gentner.
\newblock {\em Flowing waters or teeming crowds: Mental models of electricity}, pages 99--129.
\newblock Erlbaum, 1983.

\bibitem{carpenter2022science}
S.~K. Carpenter, S.~C. Pan, and A.~C. Butler.
\newblock The science of effective learning with spacing and retrieval practice.
\newblock {\em Nature Reviews Psychology}, 1(9):496--511, 2022.

\bibitem{agarwal2021retrieval}
P.~K. Agarwal, L.~D. Nunes, and J.~R. Blunt.
\newblock Retrieval practice consistently benefits student learning: A systematic review of applied research in schools and classrooms.
\newblock {\em Educational Psychology Review}, 33(4):1409--1453, 2021.

\bibitem{gjerde2020}
V.~Gjerde, B.~Holst, and S.~D. Kolst\o{}.
\newblock Retrieval practice of a hierarchical principle structure in university introductory physics: Making stronger students.
\newblock {\em Phys. Rev. Phys. Educ. Res.}, 16:013103, Apr 2020.

\bibitem{gjerde2022problem}
V.~Gjerde, V.~Havre Paulsen, B.~Holst, and S.~D. Kolst{\o}.
\newblock Problem solving in basic physics: Effective self-explanations based on four elements with support from retrieval practice.
\newblock {\em Physical Review Physics Education Research}, 18(1):010136, 2022.

\bibitem{zu2019comparing}
T.~Zu, J.~Munsell, and N.~S. Rebello.
\newblock Comparing retrieval-based practice and peer instruction in physics learning.
\newblock {\em Physical Review Physics Education Research}, 15(1):010105, 2019.

\bibitem{McCaskey2009}
T.~L. McCaskey.
\newblock {\em Comparing And Contrasting Different Methods For Probing Student Epistemology And Epistemological Development In Introductory Physics}.
\newblock PhD thesis, University of Maryland, 2009.

\bibitem{Madsen2015}
A.~Madsen, S.~B. McKagan, and E.~C. Sayre.
\newblock How physics instruction impacts students' beliefs about learning physics: A meta-analysis of 24 studies.
\newblock {\em Phys. Rev. ST Phys. Educ. Res.}, 11:010115, Jun 2015.

\end{thebibliography}

\end{document}